# High efficiency glass-based VUV metasurfaces

*Augusto Martins [ab]\*, Taylor Contreras [a], Chris Stanford [a], Mirald Tuzi [c], Justo M. Albo [c], Carlos O. Escobar [d], Adam Para [d], Alexander Kish [d], Joon-Suh Park [e], Thomas F. Krauss [a], and Roxanne Guenette [f]\*\**

[a]Department of Physics, Harvard University; Cambridge, MA 02138, United States

[b]School of Physics, Engineering and Technology University of York, York YO10 5DD, United Kingdom

[c]Instituto de Física Corpuscular (IFIC), CSIC & Universitat de València; Paterna, E-46980, Spain

[d]Fermi National Accelerator Laboratory; Batavia, IL 60510, United States

[e]John A. Paulson School of Engineering and Applied Sciences, Harvard University; Cambridge, MA 02138, United States.

[f]Department of Physics, University of Manchester; Manchester, M13 9PL, United Kingdom.

**KEYWORDS**: Vacuum ultraviolet regime, dielectric metalens, fused silica, light collection, scintillation

**ABSTRACT**: Most advances in metaoptics have been made at visible wavelengths and above; in contrast, the vacuum ultraviolet (VUV) has barely been explored despite numerous scientific and technological opportunities. Creating metaoptic elements at this short wavelength is challenging due to the scarcity of VUV transparent materials and the small sizes of the required nanostructures. Here, we present the first transmissive VUV (175 nm) metalens. By using UV-grade silica and trading-off the Nyquist requirement for subwavelength structures against feasibility of the fabrication process, we achieve a step-change in diffraction efficiencies for wavelengths shorter than 300 nm. Our large numerical aperture (NA = 0.5) metalens shows an average diffraction efficiency of (53.3 ± 1.4)%. This demonstration opens up new avenues for compact flat optic systems operating in the VUV range.

## INTRODUCTION

Metalenses are made of subwavelength scale nanostructures that modulate the phase, amplitude or polarization to focus light [1, 2]. Optical systems that typically require several elements and precisely shaped aspheric surfaces can be replaced by just one or a few planar, thin film metalenses, thus reducing device size and cost. Additionally, their fabrication is compatible with conventional semiconductor lithographic processes [3, 4], which greatly improves integration capabilities and mass manufacturability. Metalenses have successfully been demonstrated in systems that require diffraction-limited focusing [5], wide field of view imaging [6-8], achromatic focusing [9, 10], and non-imaging optics applications to increase light collection [11, 12]. However, most metalens developments have been designed for applications at wavelengths longer than 300 nm [11, 13-16]. There are two exceptions: a recently demonstrated metalens operating at a record wavelength of 50 nm [17], and a non-linear ZnO based metalens that generates and focuses 197 nm light [18], albeit both exhibiting low efficiency.

The use and detection of vacuum ultraviolet (VUV) light, of 100 nm – 200 nm wavelength, is an active area of research that finds applications in several domains such as high energy physics (HEP) [19-26], positron-electron tomography [27], photolithography [28] and biotechnology [29]. In particular, large-scale neutrino and dark matter detectors using noble elements such as argon or xenon, where scintillation light is in the VUV, would significantly benefit from very large area, highly efficient focusing elements to increase light collection at low cost and to offer enhanced discovery potential. Common optical elements in this wavelength range are either reflective or refractive bulky elements made of $CaF_2$ or $MgF_2$, which are fragile and costly materials [18]. We have identified metalenses as an exciting alternative even though they are extremely challenging to make because of material transmission losses and fabrication difficulties at these short wavelengths. For 50 nm wavelength, in the extreme ultraviolet range, the material challenge was recently addressed by exploring vacuum waveguiding in silicon nanoholes, albeit resulting in a focusing efficiency lower than 5% [17]. Additionally, the design relaxed the subwavelength requirement for the nanostructures, which limited the numerical aperture to about 0.05.

Noble elements, most commonly xenon and argon [19-26], scintillate in the 100-200 nm range and are being used in current and planned experiments for neutrino physics and dark matter searches. Detection is extremely challenging because typical detectors in the VUV range have low (~10% – 20%) efficiency [30, 31], so many experiments have resorted to wavelength shifting substances converting VUV light into the visible range that is more suitable for existing photodetectors [32]. However, the total fraction of light collected is low, which is a limiting factor for the energy threshold and resolution of high energy particle detection, crucial for rare event searches. As the scale of experiments increases, the higher number of photon detectors required becomes prohibitively expensive. The high energy scientific community is therefore highly interested in novel solutions to increase the number of photons detected while keeping the cost manageable [33, 34].

Here, we present the first experimentally realized metalens designed for the VUV (175 nm), which corresponds to the scintillation light emitted by Xenon. Our metalens is made of UV-grade pure JGS1 fused silica nanoposts etched monolithically into the substrate, which has low losses down to the VUV range. We demonstrate that our large NA design achieves the highest reported diffraction efficiency obtained for wavelengths shorter than 300 nm with an average diffraction efficiency of (53.3 ± 1.4)%. The metalens maintains similar efficiencies at other wavelengths that are interesting for other applications such as nanolithography, i.e. the 190 nm – 200 nm range. We also present, in the supporting information (SI), a highly efficient metagrating design with a higher

NA of 0.65. Simulations of the point spread function suggest that this design is capable of diffraction-limited focusing while maintaining high focusing efficiency, which suggests that this design is also suitable for imaging applications. This demonstration enables the development of thin, compact, and efficient VUV flat optic systems.

**Results and discussion**

**Metalens design**

We fabricated a metalens based on the hyperbolic phase profile where the phase control is obtained by varying the diameter of the nanoposts with a fixed center-to-center distance [1]. The focal length is 5 mm, with an NA of 0.5 (corresponding to a diameter of 5.77 mm); a higher NA (0.65) design is presented in the SI, section S1. It is known that the unit cell size of the metalens must be subwavelength to achieve high diffraction efficiency [1]. This requirement is a consequence of the Nyquist theorem [1], and it poses significant fabrication challenges at VUV wavelengths due to the very small features and high aspect ratios that are needed. We now show that the Nyquist theorem requirement can be circumvented by increasing the unit cell period and post diameters without significantly compromising efficiency. As a material, we chose UV-grade fused silica (JGS1) glass, which is extremely pure and has an extinction length of approximately 1 mm at 175 nm (see also the SI for more details). The final design is composed of 400 nm tall silica nanoposts with 160 nm unit cell size directly etched into the silica substrate, as shown in Figure 1a.

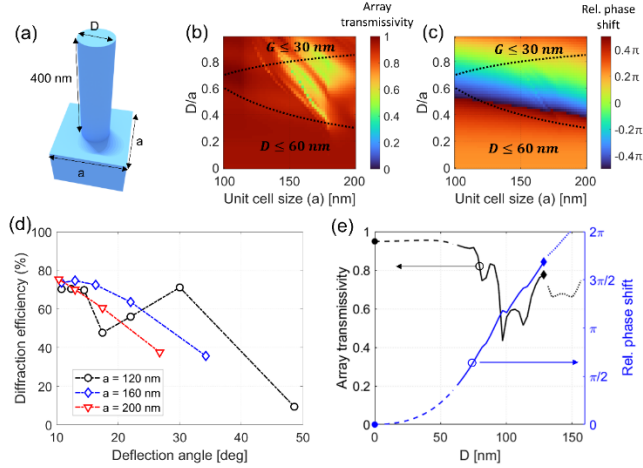

**Figure 1**. (a) Metalens unit cell design composed of 400 nm tall JGS1 rated fused silica nanoposts for operation at 175 nm. The phase control is achieved by changing the post diameter *D* for a fixed unit cell size *a*. (b) array transmissivity and (c) relative phase shift as a function of the nanopost diameter (*D*) and unit cell size (*a*), respectively. The dotted lines mark the boundaries of the fabrication limits, which correspond to $D = 60\ nm$ and $G = 30\ nm$, where *G* is the gap between the posts. The operating wavelength is 175 nm and normal incidence is used in the simulations. (d) Diffraction efficiencies of different beam steering metasurfaces (metasurface gratings) designed with unit cell sizes of 120nm, 160 nm and 200 nm following the maps in (b) and (c). (d) Array transmissivity and relative transmitted phase shift for *a* = 160 nm as function of the nanopost diameter. The dashed portion of the lines indicate the small posts (*D<60 nm*) that are removed due to the fabrication limits, and the dotted region marks the large diameter posts with narrow gaps (*G<30nm*) that were approximated by *D* = 130 *nm* (diamonds).

The simulated transmissivity and phase profile maps as a function of the post diameter and unit cell size are shown Figure 1b,c. They were computed using the rigorous coupled-wave analysis (RCWA) method [35, 36] and a glass refractive index of 1.61 at 175 nm, as listed in [37]. After iteratively optimizing the fabrication process, we were able to consistently obtain gap sizes as small as 30 nm and post diameters as small as 60 nm, as shown in Figure S2. These fabrication limits are included in Figure 1b,c, which also highlight the trade-off between diffraction efficiency and phase control. According to the Nyquist theorem [1], a high diffraction efficiency demands a half-wavelength period; decreasing the period, however, also means that the range of post-sizes that can be made within experimental constraints is reduced, hence the control over the phase profile diminishes. Considering Figure 1b, however, we note that the Nyquist requirement, surprisingly, is not as strict as expected, and that high diffraction efficiency can also be achieved for larger periods. We therefore chose a slightly larger period, which offers better phase control.

To verify this choice and to assess the diffraction performance for different unit cell periods, we simulated the beam steering property of a set of subwavelength-spaced nanopillars. The simulated supercells of nanopillars each have 120, 160 and 200 nm unit cell periodicities between the nanopillars, respectively. The beam steering metasurface, also known as a binary blazed grating, has a linear phase profile given by

$$\varphi(x) = -\frac{2\pi}{\lambda_0} x \sin(\theta),\qquad(1)$$

where $\theta$ is the deflection angle, $x$ is the horizontal coordinate on the metalens plane, and we use $\lambda_0 = 175$ nm. Figure 1d displays the simulated diffraction efficiencies, focusing only on deflection angles that yield an integer multiple of the unit cell size per supercell according to equation (1). Based on these results, we opted for a unit cell size of 160 nm since it can provide high diffraction efficiency (up to ~75%) and diffraction angles of up to 30°. A shorter period of 120 nm increases the maximum diffraction angle available by design but due to its smaller phase excursion, it produces an overall lower efficiency. Conversely, a longer period of 200 nm increases the available phase range but at the cost of a smaller maximum deflection angle (25.9°), so it also results in lower overall efficiency. For the chosen unit cell period of 160 nm, we plot the simulated phase and transmittance curves as function of post diameter in Figure 1e. The structures that lie above and below the dashed and dotted lines, respectively, were excluded from metasurface design due to fabrication constraints.

Figure 2a summarizes the fabrication process steps used to realize the metalens (see methods for more details), Figure 2b shows an optical micrograph of the metalens and Figure 2c,e show SEM micrographs of various details of the metalens. More SEM micrographs can be found in Figure S1.

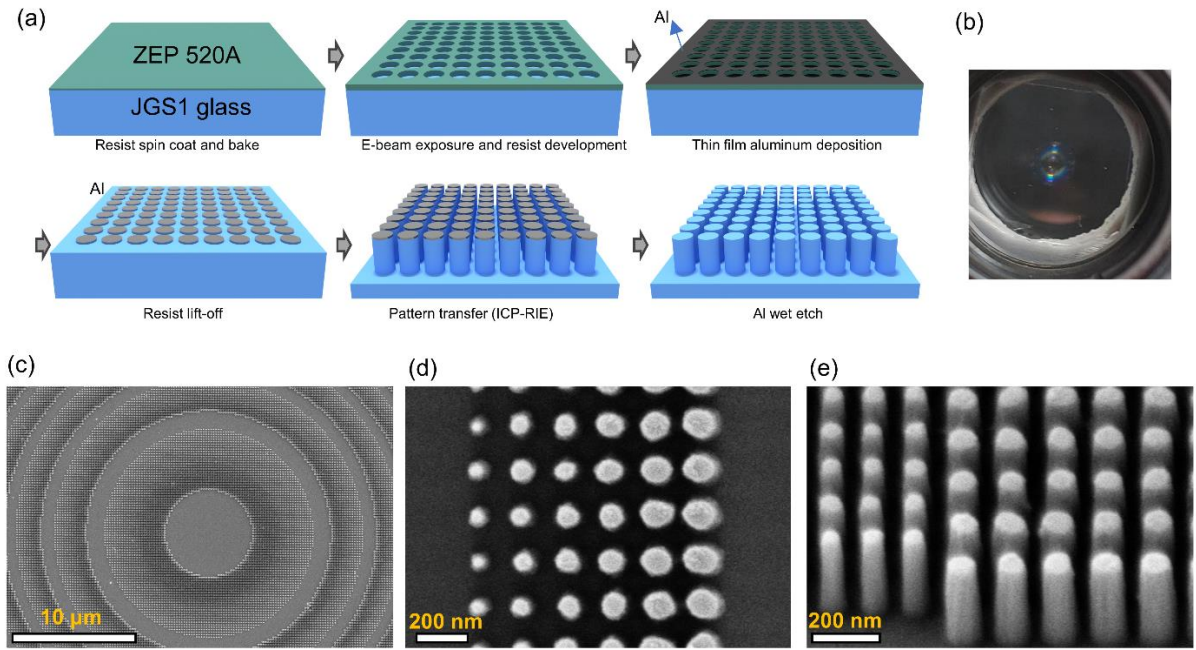

**Figure 2**. (a) Fabrication process utilized to fabricate the VUV metalens. Positive e-beam resist ZEP 520A is spin-coated on a surface of the JGS1 fused silica substrate. The metalens pattern is then written into the resist layer via e-beam exposure with an acceleration voltage of 50kV (Elionix HS-50) and developed in o-xylene. The exposed regions are removed leaving the complementary geometry of the metalens in the resist. A thin aluminum layer is then deposited by e-beam evaporation on the patterned resist. The resist is then lifted-off, leaving behind the Al that filled the exposed areas, which now has the metalens profile. Finally, the pattern is transferred to the substrate by ICP-RIE etching and the remaining Al is removed by wet etching in a 2.2% TMAH solution (b) Photo of the fabricated metalens with a focal length of 5 mm and NA = 0.5 (5.77 mm of diameter). (c), (d) and (e) show SEM micrographs of the fabricated metasurfaces.

**Metalens characterization**

We evaluated the efficiency of the metalens using an automated setup [38], as shown in Figure 3a (see Methods section for details). For this characterization, we used wavelengths of 175, 190 and 200 nm as these are technologically significant, both for high energy experiments using xenon scintillation light as well as in Deep UV lithography.

Metalens characterization typically involves the direct imaging of the point spread function (PSF) using accurate high NA and aberration free optical systems. However, the optical components needed for this measurement in the VUV regime (*e.g.*, a set of a high-NA objective lens, a corresponding tube lens designed for VUV wavelengths and a sensor) are prohibitively costly. We therefore indirectly characterize the metalens through local diffraction efficiency measurements and ensure that the light is diffracted at an angle as that matches the designed hyperbolic phase profile. In detail, we study the collected power as a function of the sensor position ($\Delta$s) and metalens beam offset ($\Delta$m) as shown in Figure 3b–d. The measured signal was normalized with respect to the power transmitted through the bare substrate to obtain the relative diffraction efficiency [39]. The black, blue and red lines in these figures mark the expected positions and angles of the -1,0 and 1 diffraction orders, respectively, according to ideal hyperbolic profile as in equation S2.7. The -1 order is responsible for the focusing of the metalens, the $0^{th}$ order preserves the incident angle and the +1 diffraction order corresponds to a diverging lens with a virtual focus.

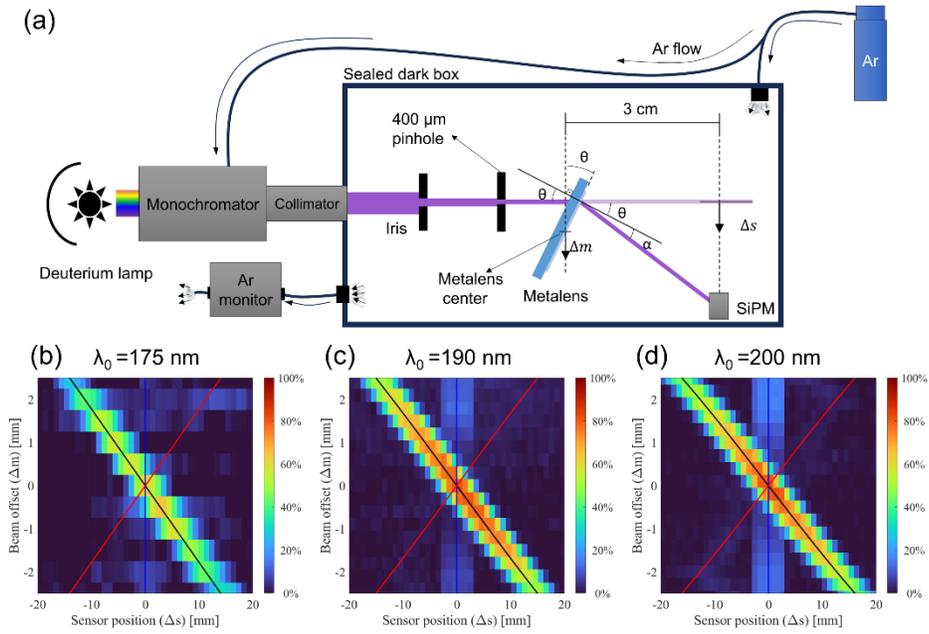

**Figure 3.** (a) Illustration of the setup used for the measurement of the metalens (image not-to-scale). The whole system was mounted inside a sealed box in an argon atmosphere to avoid VUV absorption by oxygen. To reduce the beam waist and remove spurious scattered light from the monochromator, we used an iris and a 400 μm pinhole in the optical path before the metalens. The metalens and the silicon photomultiplier sensor were mounted in translation stages that were actuated by stepper motors. The metalens rotation is actuated by a stepper motor aligned with the mount axis. The offset induced by the rotation on the metalens axis was manually corrected. For each metalens offset (Δm) we scanned the sensor in steps of 1 mm around the monochromator output position, which is also aligned with the metalens zeroth order. More details about the setup can be found in [38]. (b) – (d) relative diffraction efficiecy signal measured at 175 nm, 190 nm and 200 nm, respectively. The normalization was performed with respect to the measurement through bare substrate. The red, black and blue lines mark the expected positions by the -1,0 and 1 diffracted order according to equation (S2.6).

Figure 4a–c shows a comparison of both measured and simulated diffraction efficiencies for wavelengths of 175 nm, 190 nm, and 200 nm, with respect to the positions of the metalenses and for the 0 and -1 deflection orders. These curves were derived from the peak values presented in Figure 3b–d. Solid lines represent the experimentally measured diffraction efficiencies, while dashed lines indicate efficiencies predicted by the simulation technique detailed in [40] The efficiency at all wavelengths decreases with the distance to the metalens center (equivalently when the deflected angle increases), due to lower Nyquist sampling. Near the center of the metalens, the measured diffraction efficiencies are as high as 65%, 77% and 70% when operating at 175 nm, 190 nm and 200 nm, respectively. The average diffraction efficiency across different sections of the metalens is found to be $(53.3 \pm 1.4)$ %, $(63.9 \pm 2.0)$% and $(65.0 \pm 0.2)$% when operating at 175 nm, 190 nm, and 200 nm, respectively. These results are the first of a kind in the VUV range and, remarkably, the measured efficiencies set new benchmarks for performance at wavelengths below 300 nm (see Table S2 for a comprehensive summary of the efficiencies at varying wavelengths reported in literature).

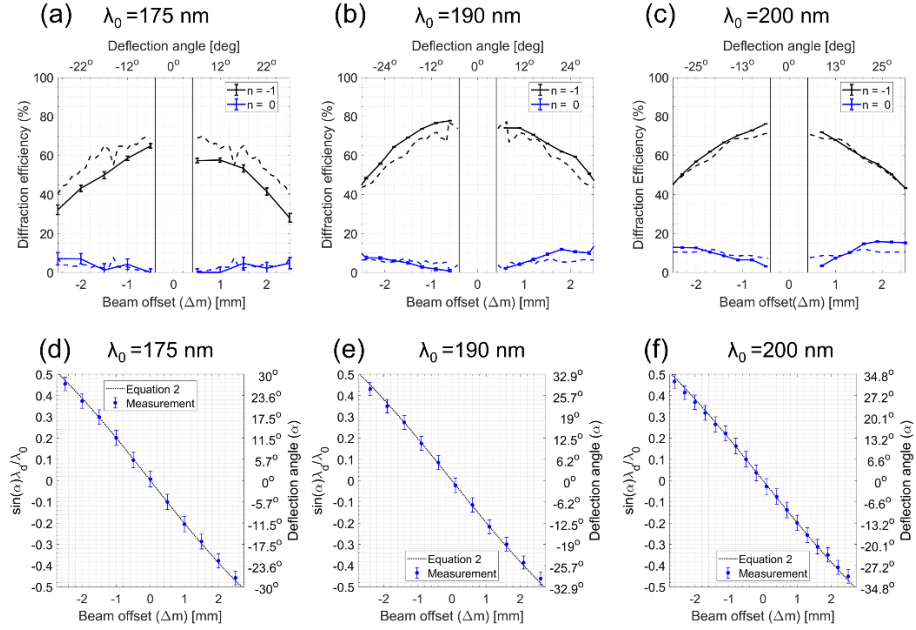

**Figure 4.** (a) – (c) Measured (solid lines) diffraction efficiency extracted from Figure 3. The black and blue lines show, respectively, the diffraction efficiency of the n = -1 (focusing) and n = 0 (zeroth) orders. The dashed lines show the simulated diffraction efficiencies. The vertical lines enclose the region where the orders overlap. (b) – (d) measured and expected deflection angles, $\alpha$, as function of the beam offset at 175 nm, 190 nm and 200 nm.

The measured diffraction angles are in good agreement with the values predicted by equation S2.6, which is a strong indication that the metalens is properly modulating the hyperbolic phase profile. At normal incidence, the diffracted angle, $\alpha$, by a hyperbolic phase profile is given by, according to equation S2.6

$$\sin\alpha\,\frac{\lambda_d}{\lambda_0}=-\frac{\Delta m}{\sqrt{\Delta m^2+f^2}}\equiv\frac{\lambda_d}{2\pi}\frac{\partial\varphi(r)}{\partial r}\bigg|_{r=\Delta m}\,,\qquad(2)$$

where $\lambda_d$ and $\lambda_0$ are the design and operating wavelengths, respectively, and $\varphi(r)$ the hyperbolic phase profile, given by equation S2.3. The wavelength ratio on the left side of equation (2) appears naturally from the generalized law of refraction when the operating wavelength is different from the design wavelength [41] and it is manifested as chromatic aberration in the diffracted ray angle $\alpha$. However, equation (2) shows that $\sin \alpha \frac{\lambda_d}{\lambda_0}$ should be ideally identical to the hyperbolic phase profile gradient, which is given by the right side of equation (2), and wavelength independent. To check if our realized metalens preserves this quantity, we plotted it as function of the beam offset when operating at 175nm, 190 nm and 200 nm in Figure 4d–f, respectively, over a plot of the ideal hyperbolic profile gradient as given by the middle term of equation (2). The measured diffracted angles were calculated with respect to the peak positions in the power map measurements in Figure 3b-d. The right axis of Figure 4d–f shows the corresponding deflected angles $\alpha$. For all measured wavelengths, the metalens preserved the hyperbolic phase profile gradient satisfactorily. This measurement indicates that each Fresnel zone bends the ray towards the focal point at 175 nm, as expected from the hyperbolic profile [42]. However, at the other operating wavelengths, although the hyperbolic phase gradient is preserved, chromatic aberration, manifested as the design to operating wavelength ratio, changes the deflected angle (compare the right axes on Figure 4d–f), and the rays no longer go exactly to the focal point. For operation at these wavelengths, the phase profile should be properly adjusted to avoid chromatic aberration using methods such as structurally-induced dispersion-engineering [10].

For light collection applications, it is also interesting to assess the performance of the concentrator at oblique incidence. Figure 5a–d show the measured (solid) and simulated (dashed) diffraction efficiencies at 7.5°, 15°, 22.5° and 30° of incidence at 175 nm. The efficiency curves are no longer symmetric at oblique incidence because the metalens gradient has different signs on each side, which is necessary for focusing light. At positive offsets, the metalens gradient and the incoming light in-plane wave vector ($k_{\parallel}$) have the same direction, according to our angle definition and Figure 3a. Therefore, the transmitted beam's in-plane wave vector has much higher absolute values, likely leading to coupling with higher-order modes and reducing the amount of light going into this order [43]. Figure 5a–d show the focusing order efficiency sharply decreasing at positive positions for steeper angles of incidence. On the contrary, at negative positions, the metalens gradient is positive and is subtracted from the incoming in plane wave vector, which minimizes the scattering to higher orders. Consequently, the efficiency on this side is more tolerant to oblique incidence. Based on the symmetry, the same analysis can be made if the angles of incidence were positive, and the x-axis reversed. Overall, our metalens offers average efficiencies of $(52.6 \pm 1.7)\%$, $(45.8 \pm 1.9)\%$, $(30.9 \pm 2.0)\%$ and $(17.36 \pm 2.0)\%$ for incidence angles of 7.5°, 15°, 22.5° and 30° respectively, which provides interesting application opportunities for light collection applications.

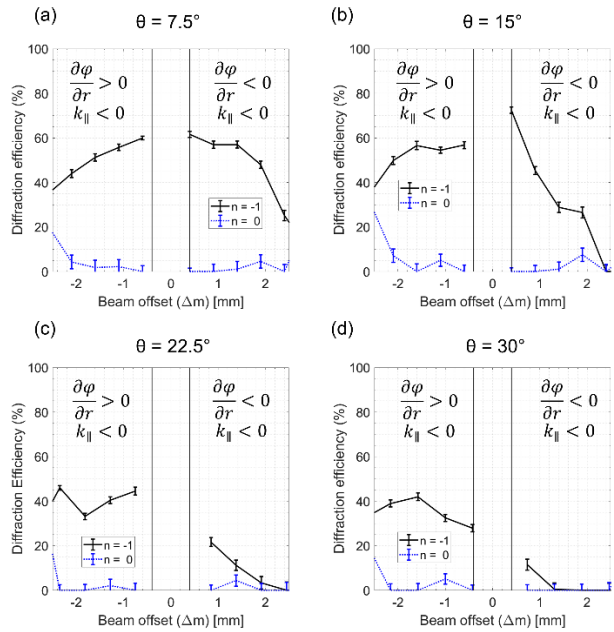

**Figure 5.** (a) – (d) VUV diffraction efficiency as a function of the beam offset position (Δm) at 7.5°, 15°, 22.5° and 30° angles of incidence, respectively. The beam waist is approximately 400 μm and the operating wavelength is 175 nm. The black and blue dotted lines show, respectively, the diffraction efficiency of the n = -1 (focusing) and n = 0 (zeroth) orders. According to our angle

definition in Figure 3a, the metalens gradient is positive in the region Δm<0 and negative in the region Δm>0. The in-plane momentum of the incoming ray, $k_{\parallel}$, is, according to Figure 3a, smaller than 0.

**Optical performance simulation**

This design could also be used for imaging applications, but our measurement is not accurate enough to capture small wavefront deviations among the Fresnel zones that could impact the point spread function (PSF). However, the combination of high efficiency and the strong agreement between the measured and experimental hyperbolic phase gradients, derived from the Fresnel zone measurements, along with the high fabrication error tolerance of glass-based metalenses [4], suggest that our VUV metalens has potential for imaging capabilities. Given the elevated challenges and costs involved in realizing precise optical measurements in the VUV range, we resorted to fully vectorial wave simulations to assess the VUV metalens point spread function (PSF). Our simulation accounts for the nanopost design by rigorously solving smaller regions of the metasurface on RCWA and patching the transmitted full field [40]. Propagation in free space is performed using vectorial angular spectrum formalism [44]. We simulated the PSF at normal incidence of a smaller diameter metalens with the same NA of 0.5 and focal length $f$ =1 mm operating at 175 nm. Such simulation approach using smaller apertures but with the same NA has been previously used to study the focusing properties of glass-based metalenses operating at 632 nm with very good agreement with experiments on larger metalenses [3]. Propagation loss through the substrate is neglected in this simulation. Figure 6a shows the longitudinal cross section along the optical of the focusing profile by the metalens around the focal plane, which shows a highly symmetric focusing profile expected for the hyperbolic profile. Figure 6b shows the PSF at the focal plane with the Airy disk function, both normalized with respect to the Airy disk intensity and amplitude. The metalens simulated PSF full width at half maximum is 183.4 nm, close to the 175 nm diffraction limit, and with a Strehl ratio of 0.96, both suggesting that this design can focus on the diffraction limit despite the design constraints we imposed to facilitate fabrication. These observations are corroborated by comparing the metalens modulation transfer function (MTF) with the corresponding diffraction limited MTF in Figure 6c. The focusing efficiency, calculated on an aperture three times larger than the Airy disk FHWM[4], is 68.6%, which is about 15% larger than the average measured Fresnel zone efficiency.

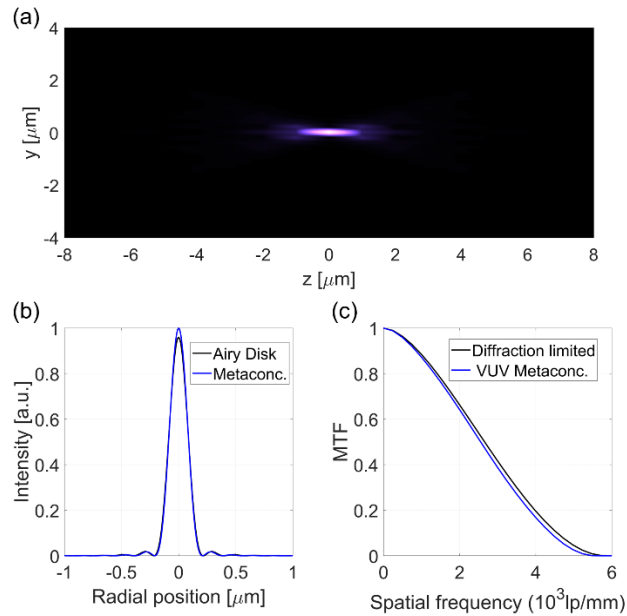

**Figure 6.** (a) simulated metalens with $f$=1mm and NA = 0.5 longitudinal focusing intensity distribution along the optical axis. (b) simulated metalens PSF with the ideal Airy disk with an NA = 0.5. (c) simulated modulation transfer function (MTF) of the VUV metalens and of the diffraction limited Airy disk. The operating wavelength is 175 nm.

## Conclusions

We have demonstrated the first highly efficient, high numerical aperture metalens operating in the VUV range. The metalens is realized in JGS1 rated fused silica glass nanoposts, which are transmissive in the VUV range, and are polarization insensitive. The metalens is designed to operate at 175 nm and has a focal length of 5 mm with a numerical aperture of 0.5 reaching a high average diffraction efficiency of $(53.3 \pm 1.4)\%$. The metalens can reach up to 65% diffraction efficiency for smaller deflection angles (smaller NAs) and maintains high efficiencies for incidence angles as large as 30°, expanding the range of potential applications. We also report that the numerical aperture can be further increased, while preserving high efficiency, by adopting a metagrating design. High transmission efficiency is achieved by exploiting the trade-off between the sub-wavelength period requirement (Nyquist) and available phase range excursion where the posts diameter and gaps are limited by fabrication. Simulations suggest that this design could also potentially be used for imaging applications in the VUV range with diffraction-limited performance. This approach results in the highest transmission efficiency metalens for wavelength below 300 nm. The metalens is produced using a conventional e-beam lithography fabrication process, which is shown to be repeatable, thus the design is compatible with mass manufacturing CMOS processes based on DUV immersion lithography [3, 4, 45, 46], allowing for large-scale fabrication. Our VUV metalens paves the way to applications that could be transformed by the metaoptics paradigm, such as future large-scale particle detectors using VUV scintillation light to study neutrinos and dark matter, as well as new technologies for positron-electron tomography imaging.

## Methods

**Metalens fabrication.** The metalenses were patterned on ZEP 520A (Zeon SMI) positive resist using e-beam lithography and then transferred to an aluminum mask with a lift-off process. The pattern was transferred to the glass substrate using an optimized ICP-RIE vertical etching recipe[4]. The remaining aluminum layer was finally wet etched in a 2.2% TMAH solution for about 10 min.

**Metalens characterization setup.** We employed an unpolarized VUV beam generated from a vacuum monochromator (McPherson, 234/302) fitted with a deuterium lamp. The monochromator featured with a mirror-based compact vacuum collimating chamber (McPherson, 100-103134), and an iris paired with a 400 μm pinhole to enhance beam collimation and and to spatially filter the diffraction from the iris, which results in a collimated beam with a diameter of 400 μm. The metalens was mounted on a motorized stage with translational and rotational degrees of freedom, allowing for a scan and rotation of the lens with respect to the beam. The sensor, a VUV-sensitive silicon photomultiplier (Hamamatsu, VUV4), was mounted on a linear stage placed 3 cm away from the metalens. The sensor was moved laterally to collect the diffracted light (see Figure 3a). The whole system was mounted inside a sealed box in an argon atmosphere to avoid VUV absorption by oxygen [47].

ASSOCIATED CONTENT

**Supporting Information**.

JGS1 extinction coefficient, additional SEM micrographs, VUV metagrating design, expected position of the diffraction lines, efficiency of UV metasurfaces previously demonstrated (PDF)

AUTHOR INFORMATION

**Corresponding Author**

*augusto.martins@york.ac.uk

**roxanne.guenette@manchester.ac.uk

## Author Contributions

The manuscript was written through contributions of all authors. All authors have given approval to the final version of the manuscript.

## Funding Sources

Sloan Foundation under a 2021 Alfred P. Sloan Research Fellowship (AM, CS)

European research council grant ERC-2020-SyG-951281(RG, TC and CS)

Department of Energy (DOE) LDRD program L2021.011 (COE, AK, AP)

## ACKNOWLEDGMENT

Part of this research and support for A. Martins and C. Stanford was funded by the Sloan Foundation under a 2021 Alfred P. Sloan Research Fellowship. This document was prepared by using in part the resources of the Fermi National Accelerator Laboratory (Fermilab) and the Noble Liquid Test Facility (NLTF), a U.S. Department of Energy, Office of Science, Office of High Energy Physics HEP User Facility. Fermilab is managed by Fermi Research Alliance, LLC (FRA), acting under Contract No. DE-AC02-07CH11359. The authors R.G., T.C. and C.S. acknowledge financial support by the European research council (Grant ERC-2020-SyG-951281). C.O.E, A.K. and A.O. acknowledge financial support by the Department of Energy (DOE) LDRD program L2021.011. This work was performed in part at the Harvard University Center for Nanoscale Systems (CNS); a member of the National Nanotechnology Coordinated Infrastructure Network (NNCI), which is supported by the National Science Foundation under NSF award no. ECCS-2025158. Most computations in this paper were run on the FASRC cluster supported by the FAS Division of Science Research Computing Group at Harvard University.

ABBREVIATIONS

VUV vacuum ultraviolet; NA numerical aperture; PSF point spread function; RCWA rigorous coupled wave analysis.

# Supporting information for: High efficiency glass-based VUV metasurfaces

*Augusto Martins [ab]\*, Taylor Contreras [a], Chris Stanford [a], Mirald Tuzi [c], Justo M. Albo [c], Carlos O. Escobar [d], Adam Para [d], Alexander Kish [d], Joon-Suh Park [e], Thomas F. Krauss [a], and Roxanne Guenette [f]\*\**

[a]Department of Physics, Harvard University; Cambridge, MA 02138, United States

[b]School of Physics, Engineering and Technology University of York, York YO10 5DD, United Kingdom

[c]Instituto de Física Corpuscular (IFIC), CSIC & Universitat de València; Paterna, E-46980, Spain

[d]Fermi National Accelerator Laboratory; Batavia, IL 60510, United States

[e]John A. Paulson School of Engineering and Applied Sciences, Harvard University; Cambridge, MA 02138, United States.

[f]Department of Physics, University of Manchester; Manchester, M13 9PL, United Kingdom.

**Corresponding Author**

\*augusto.martins@york.ac.uk

\*\*roxanne.guenette@manchester.ac.uk

**Supporting information**

**Section S1 – JGS1 extinction coefficient the VUV range**

Figure S1

Table S1

**Section S2 – Micrographs**

Figure S2

**Section S3 – Metagrating based VUV metalens**

Figure S3

Figure S4

**Section S4 – Expected positions of the diffraction lines**

**Section S5 – Experimentally demonstrated UV metasurfaces in the literature**

Table S2

**Section S1 – JGS1 extinction coefficient the VUV range**

The metalenses and meta-gratings were fabricated on JGS1 ultraviolet (UV) grade silica glasses which are free of bubbles and transparent in the ultraviolet and visible regions. We measured the transmittance spectrum of the substrate from the VUV range up to the UVA range. For wavelengths shorter than 200 nm, we used the setup shown in Figure 3a, purging high-purity argon gas throughout the measurement. The transmission through the bare substrate was measured and normalized with the incoming power. For wavelengths longer than 200 nm we used a Thorlabs SLS204 deuterium light source for 200 nm - 700 nm coupled to a DUV-rated M114L01 fiber. We collected and collimated the fiber output light using a NA=0.4 Edmund Optics reflective microscope objective lens with an iris. We then measured the transmitted spectrum using a thermoelectrically cooled ASEQ LR1 spectrometer. The measured UV transmission spectrum of a $d = 500$ μm ± 50 μm thick JGS1 substrate, which is the same used to fabricate the metalens, is shown in Fig S1. It is transparent in most of the spectrum and still capable of transmitting over 35% at the wavelength of 165 nm.

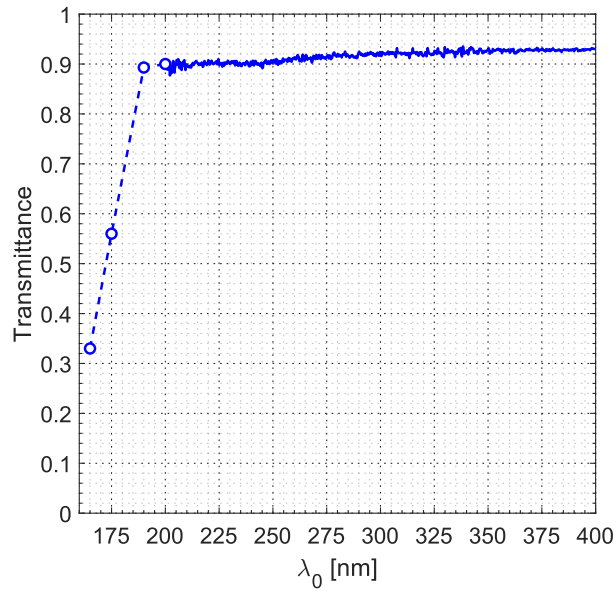

**Figure S1.** Transmittance spectrum of the JGS1 glass substrate.

Measurement of the transmittance spectrum of a 500 μm thick JGS1 substrate throughout the UV range. For the measurements at 165 nm, 175 nm, 190 nm and 200 nm, we used the setup shown in Figure 3a, where the substrate was illuminated with a short wavelength band from the monochromator. For wavelengths longer than 200 nm we used a Thorlabs deuterium SLS204 Deuterium broadband light source (200 nm - 700 nm) coupled to a DUV rated M114L01 fiber to illuminate the metalens coupled. We collected and collimated the fiber output light using an NA=0.4 Edmund Optics reflective microscope objective lens with an iris. We then measured the transmitted spectrum using a thermoelectrically cooled ASEQ LR1 spectrometer.

The JGS1 transmission in the VUV range was then used to estimate its extinction length. The transmission equation for a slab in embed in air under incoherent light illumination is given by [1]

$$T = \frac{te^{-\alpha d}}{1 - re^{-2\alpha d}},  \tag{M1.1}$$

where $T$ is the measured transmission coefficient, $d$ is the substrate thickness, $\alpha$ is the extinction coefficient and $t$ ($r$) is the combined transmission (reflection) coefficient between the interfaces, given by

$$t = \left(\frac{2}{n+1}\right)^4 n^2,  \tag{M1.2}$$

$$r = \left(\frac{n-1}{n+1}\right)^4,  \tag{M1.3}$$

where $n$ is the glass refractive index [2]. Solving (M1.1), we find that the extinction length $l = \alpha^{-1}$ is given by

$$l = \frac{d}{\ln \dfrac{2Tr}{\sqrt{t^2 + 4T^2 r}}}, \qquad \text{(M1.4)}$$

The transmission coefficients in the VUV range and the expected extinction lengths are listed in Table S1. The extinction length is larger than 500 μm for all wavelengths, which makes this material very suitable for VUV metalenses since the required post thickness is on the order of a few hundred nanometers. To minimize the substrate losses, one could use thinner substrates, or nanoholes dug into a glass membrane [3, 4].

**Table S1.** Transmittance measurement of a 500μm±50μm thick JGS1 rated silica glass substrate at different wavelengths. From equation (S1) we inferred the extinction lengths assuming a refractive index given by [2].

| Wavelength | 165 nm | 175 nm | 190 nm | 200 nm |
|---|---|---|---|---|
| Transmittance (%) | 33±1 | 56 ± 1 | 89 ± 0.1 | 91±0.05 |
| Extinction length $l$ (mm) | 0.50±0.07 | 1.07±0.06 | 33.1±2.9 | 417.7±19.4 |

**Section S2 – Micrographs**

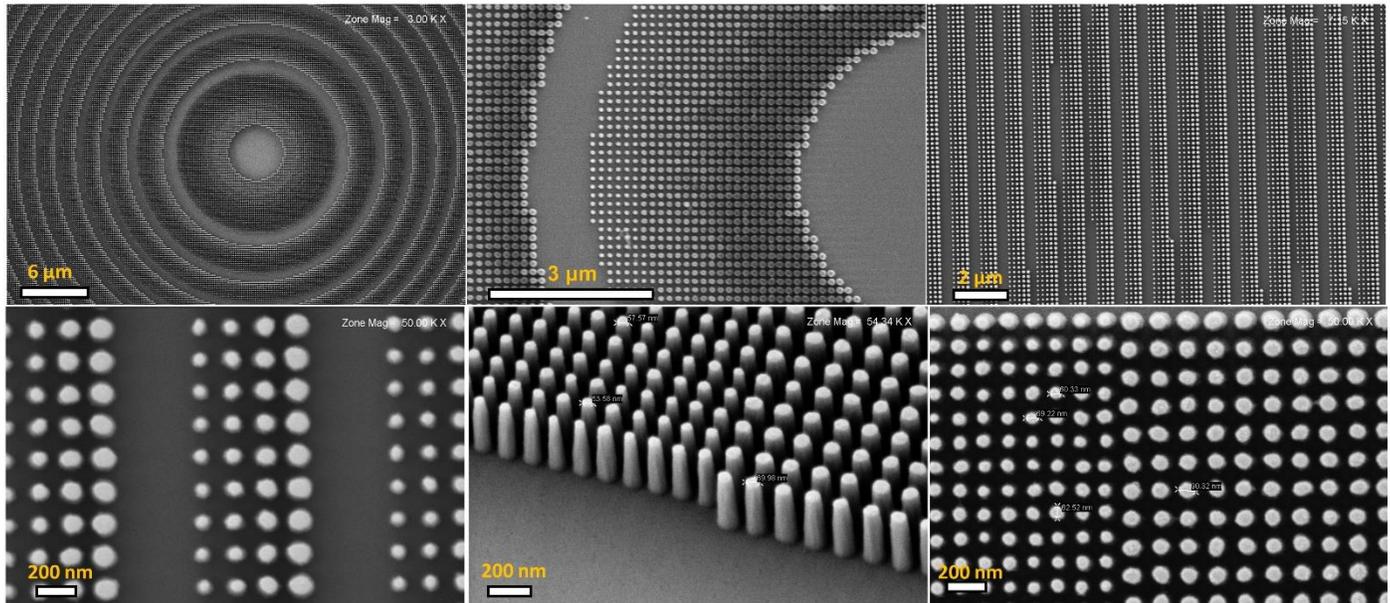

**Figure S2**. SEMs of fabricated VUV metalens. Our process allowed us to consistently obtain features as small as 60 nm and gaps of approximately 30nm.

**Section S3 – Metagrating based VUV metalens**

The efficiency of the proposed metasurface design decreases for high deflection angles, as shown in Figure 3a of the main text. One approach that has been explored in the literature to address this issue is the use of metagratings in the high deflection region [5-8]. Metagratings are structures that are optimized to deflect light at a given angle using the supercell concept. In other words, the structure geometry is optimized to enhance the first diffraction order efficiency. This enhancement can be achieved by optimizing the geometry of a finite number of structures [6, 8] or by generating free-form structures using topology optimization algorithms [5, 7]. Here, we used the former approach and optimized the geometry of a pair of 400 nm rectangular nanoposts as shown in the inset of Figure S3. We chose the supercell period $P$ according to the diffraction equation, that is

$$P = \frac{\lambda_0}{\sin \theta}, \qquad (S1.1)$$

where, $\lambda_0 = 175$ nm is the operating wavelength and $\theta$ the deflection angle. We fixed the period on the orthogonal direction at 200 nm. We used a generic algorithm optimization method on an rigorous coupled wave analysis code to optimize the dimension of the posts and the gap between them (see the inset of Figure S3a). Figure S3a shows the efficiency of the optimized structures and the efficiency of steering metasurfaces obtained with the phase map approach shown in Figure 3c. Note that the metagrating approach has a higher efficiency for angles larger than 20° than the phase map design. Thus, we combined both the phase map and the metagrating designs to fabricate an NA=0.65 metalens. In this metalens, the region that deflects light with angles smaller than 20° is based on the phase map design, and the metagrating design is applied on the remaining part of the metalens. To test this design, we fabricated the central stripe of an NA = 0.65 metalens with a focal length of 9 mm. Note that the largest deflection angle by this metalens is around 40°. The total dimensions of the stripe are 1.7mm × 15.39mm. A photo of the metalens is shown in Fig. S3 along with some SEM images of some optimized metagratings. Figure. S3b-c show the measured diffraction efficiency of the metalens as function of the beam offset operating at 175 nm, 190 nm, and 200 nm. Note that the onset of the optimized structure can be easily distinguished by the sharp discontinuity at around Δm = ± 4 mm, which clearly indicates the higher performance of the optimized metagrating. The average diffraction efficiencies of the metalens operating at 175 nm, 190 nm and 200nm are $(63.68 \pm 1.2)\%$, $(47.52 \pm 0.3)\%$ and $(50.24 \pm 0.4)\%$, respectively.

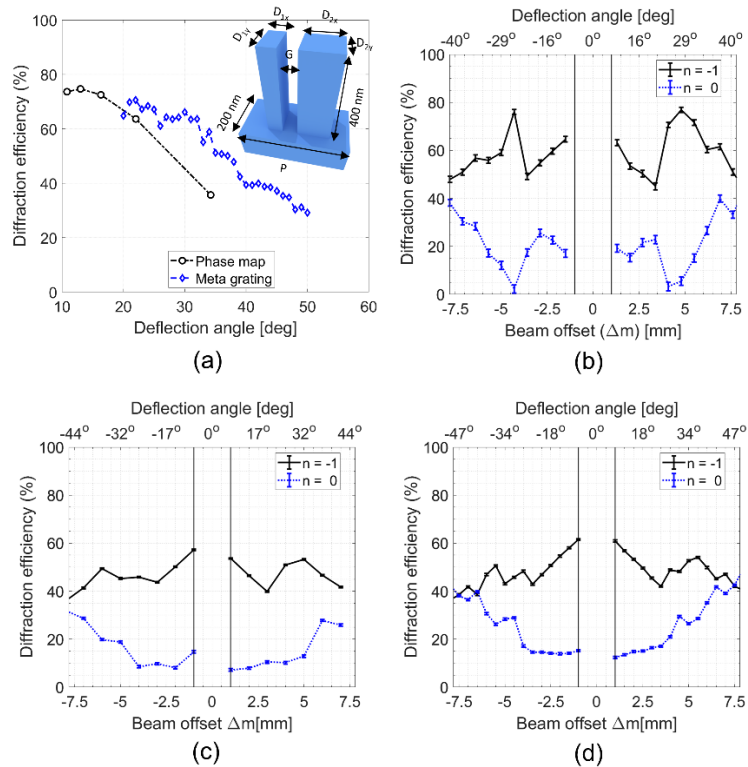

**Figure S3.** (a) Design (inset) and simulated diffraction efficiency of the optimized VUV metagrating (blue diamonds). SEMs of the fabricated metagratings are shown in Fig. S4. For comparison, the black dots show the simulated diffraction efficiency of metagrating designed using the phase map of Fig. 3C. The operating wavelength is 175 nm in the simulations. (b) – (d) Measured diffraction efficiency of the metagrating based VUV metalens operating at 175 nm, 190 nm and 200 nm, respectively.

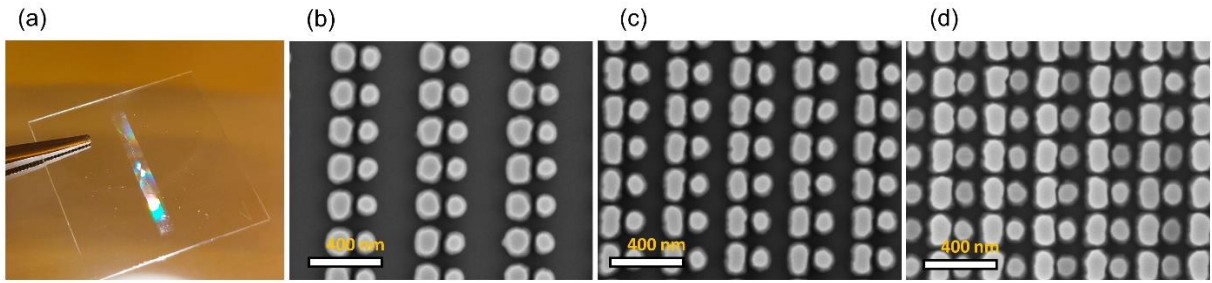

**Figure S4.** (a) Photo of the metagrating-based VUV metalens stripe. The metalens has an NA = 0.65 with a focal length of 9 mm. The total dimensions of the stripe are 1.7mm × 15.39mm. The design wavelength is 175 nm. (b)-(d) SEM micrograph of VUV metagratings optimized to deflect the beam at 22°, 32°, and 40°, respectively.

**Section S4 – Expected positions of the diffraction lines**

The expected sensor position ($\Delta s$) for a given beam offset ($\Delta m$) and angle of incidence is shown in Fig. 3A of the main text. Simple trigonometry gives us the following relation

$$\Delta s = \left(d + \Delta m \tan\theta\right)\tan\left(\arcsin\left(\theta + \alpha\right)\right), \tag{S2.1}$$

where $d = 3$cm is the distance from the metalens center to the sensor, $\theta$ is the angle of incidence and $\alpha$ the diffracted angle by the metalens (see Fig. 3A). In the first order approximation, the metalens behaves as a binary blazed grating with supercell period defined by the spatial frequency of the metalens [9]. That is

$$P = \frac{2\pi}{\left|\dfrac{\partial\varphi}{\partial r}\right|_{r=\Delta m}}, \tag{S2.2}$$

where $\varphi$ is the metalens phase profile, which is given by the hyperbolic phase profile

$$\varphi(r) = -\frac{2\pi}{\lambda_d}\left(\sqrt{r^2 + f^2} - f\right), \tag{S2.3}$$

where $r$ is the radial distance of the metalens, $f$ the focal length, and $\lambda_d = 175$nm is the design wavelength. From the grating equation, it follows that

$$\frac{2\pi}{\lambda_0}\sin\alpha = n\frac{2\pi}{P} - \frac{2\pi}{\lambda_0}\sin\theta, \tag{S2.4}$$

where $\lambda_0$ is the operating wavelength, and $n$ is the diffraction order. We assumed unitary refractive index for the incoming and transmitted media. The negative sign in the incidence term in equation S2.3 comes from the angle and axis definitions in Fig. 3A. Thus, from equations S2.2 and S2.4:

$$\sin\alpha = \lambda_0 n\left|\frac{\partial\varphi}{\partial r}\right|_{r=\Delta m} - \sin\theta, \tag{S2.5}$$

Equation (S2.4) is commonly known as the generalized law of refraction when $n = -1$ [10]. From equations (S2.5) and (S2.3), we can calculate the diffracted angle by the metalens as

$$\sin \alpha = n \frac{\Delta m}{\sqrt{\Delta m^2 + f^2}} \frac{\lambda_0}{\lambda_d} - \sin \theta \qquad \text{(S2.6)}$$

From equations (S2.1) and (S2.6) it follows that the expected sensor position ($\Delta s$) for a given beam offset and angle of incidence is given by

$$\Delta s = \left( d + \Delta m \tan \theta \right) \tan \left( \arcsin \left( n \frac{\Delta m}{\sqrt{\Delta m^2 + f^2}} \frac{\lambda_0}{\lambda_d} - \sin \theta \right) + \theta \right) \qquad \text{(S2.7)}$$

**Section S5 – Experimentally demonstrated UV metasurfaces in the literature**

**Table S2.** Experimentally demonstrated UV metasurfaces

| Range | Wavelength | Material | NA | Polarization | Efficiency |
|---|---|---|---|---|---|
| EUV | 50 nm | Crystalline silicon membrane [4] | 0.05 | Unpolarized | 4.8% |
| VUV | 175 nm | Silica (this work) | 0.5 | Unpolarized | $(53.33 \pm 1.42)\%$ |
| | | | 0.65 | | $(63.68 \pm 1.19)\%$ |
| | 190 nm | | 0.5 | | $(63.92 \pm 1.95)\%$ |
| | | | 0.65 | | $(47.52 \pm 0.31)\%$ |
| | 200 nm | | 0.5 | | $(65.03 \pm 0.15)\%$ |
| | | | 0.65 | | $(50.24 \pm 0.4)\%$ |
| DUV | 266 nm | Hafnium oxide [11] | n.a. (holograms) | Circular polarization (PB-phase) | $(60.67 \pm 2.60)\%$ |
| UVA | 325 nm | | 0.6 | | $(55.17 \pm 2.56)\%$ |
| | 364 nm | | 0.6 | | $(56.28 \pm 1.37)\%$ |
| | 355 nm | Niobium pentoxide [12] | n.a. (holograms) | Circular polarization (PB-phase) | 79.6% |
| | 360 nm | Silica and polymer [8] | 0.9 | Unpolarized | 13.9% |
| | 380 nm | Crystalline silicon [13] | n.a. (holograms) | Circular polarization (PB-phase) | 15% |